\documentclass
[preprint,showpacs,preprintnumbers,amsmath,amssymb,superscriptaddress]{revtex4}

\usepackage{graphicx}% Include figure files
\usepackage{dcolumn}% Align table columns on decimal point
\usepackage{bm}% bold math
\usepackage{color}

\begin{document}

%\preprint{PREPRINT (\today)}

%\newpage

\title{Spectromicroscopy of electronic phase separation in K$_x$Fe$_{2-y}$Se$_2$ superconductor}

\author{M. Bendele}
\affiliation{Dipartimento di Fisica, Universit\'a di Roma ``La Sapienza'' - P. le Aldo Moro 2, 00185 Roma, Italy}

\author{A. Barinov}
\affiliation{Sincrotrone Trieste S.C.p.A., Area Science Park, 34012 Basovizza, Trieste, Italy}

\author{B. Joseph}
\affiliation{Dipartimento di Fisica, Universit\'a di Roma ``La Sapienza'' - P. le Aldo Moro 2, 00185 Roma, Italy}

\author{D. Innocenti}
\affiliation{Rome International Center for Materials Science Superstripes (RICMASS), Via dei Sabelli 119A, 00185 Roma, Italy}

\author{A. Iadecola}
\affiliation{Sincrotrone Trieste S.C.p.A., Area Science Park, 34012 Basovizza, Trieste, Italy}

\author{A. Bianconi}
\affiliation{Rome International Center for Materials Science Superstripes (RICMASS), Via dei Sabelli 119A, 00185 Roma, Italy}

\author{H. Takeya}
\affiliation{National Institute for Materials Science, 1-2-1 Sengen, Tsukuba 305-0047, Japan}

\author{Y. Mizuguchi}
\affiliation{National Institute for Materials Science, 1-2-1 Sengen, Tsukuba 305-0047, Japan}

\author{Y. Takano}
\affiliation{National Institute for Materials Science, 1-2-1 Sengen, Tsukuba 305-0047, Japan}

\author{T. Noji}
\affiliation{Department of Applied Physics, Tohoku University, 6-6-05 Aoba, Aramaki, Aoba-ku, Sendai 980-8579, Japan}

\author{T. Hatakeda}
\affiliation{Department of Applied Physics, Tohoku University, 6-6-05 Aoba, Aramaki, Aoba-ku, Sendai 980-8579, Japan}

\author{Y. Koike}
\affiliation{Department of Applied Physics, Tohoku University, 6-6-05 Aoba, Aramaki, Aoba-ku, Sendai 980-8579, Japan}

\author{M. Horio}
\affiliation{Department of Physics, University of Tokyo, 7-3-1 Hongo, Tokyo 113-0033, Japan}

\author{A. Fujimori}
\affiliation{Department of Physics, University of Tokyo, 7-3-1 Hongo, Tokyo 113-0033, Japan}

\author{D. Ootsuki}
\affiliation{Department of Physics, University of Tokyo, 5-1-5 Kashiwanoha, Chiba 277-8561, Japan}

\author{T. Mizokawa}
\affiliation{Department of Complexity Science and Engineering, University of Tokyo,5-1-5 Kashiwanoha, Chiba 277-8561, Japan}
\affiliation{Department of Physics, University of Tokyo, 5-1-5 Kashiwanoha, Chiba 277-8561, Japan}
 
\author{N. L. Saini}
%\email{Naurang.Saini@roma1.infn.it}
\affiliation{Dipartimento di Fisica, Universit\'a di Roma ``La Sapienza'' - P. le Aldo Moro 2, 00185 Roma, Italy}

%*Correspondence to [Naurang.Saini@roma1.infn.it]

\begin{abstract}
\textbf{Structural phase separation in \emph{A}$_{x}$Fe$_{2-y}$Se$_2$ system has been studied by different experimental techniques, however, it should be important to know how the electronic uniformity is influenced, on which length scale  the electronic phases coexist, and what is their  spatial distribution. Here, we have used novel scanning photoelectron microscopy (SPEM) to study the electronic phase separation in K$_{x}$Fe$_{2-y}$Se$_2$, providing a direct measurement of the topological spatial distribution of the different electronic phases. The SPEM results reveal a peculiar interconnected conducting filamentary phase that is embedded in the insulating texture. The filamentary structure with a particular topological geometry could be important for the high T$_c$ superconductivity in the presence of a phase with a large magnetic moment in \emph{A}$_{x}$Fe$_{2-y}$Se$_2$ materials.}
\end{abstract}

%\pacs{0 0 0}

\maketitle

The discovery of superconductivity in Fe-based  materials \cite{Kamihara08} gave a new boost to the research on this intriguing quantum phenomenon. Among others, the FeSe was found to show superconductivity at a \emph{T}$_{\rm c}\sim$ 8 K \cite{Hsu08}, that could be increased to over 30 K by physical pressure \cite{Medvedev_Nat_09}. Furthermore, it can be increased to similar values by intercalating alkali atoms \emph{A}, yielding in a new series of \emph{A}$_{x}$Fe$_{2-y}$Se$_2$ materials \cite{Guo10,Mizuguchi11}. One of the interesting features of $A_{x}$Fe$_{2-y}$Se$_2$ system is the intrinsic phase separation in coexisting crystallographic phases \cite{Liu11a,Wang11,Li11b,Ricci11b,Shermadini12,Chen_PRX,Yuan12,Wen12}. The minority phase has a stoichiometry close to $A_x$Fe$_2$Se$_2$ (122) \cite{Fang11}, embedded into a majority phase that manifests iron-vacancy ordering.  This majority phase has a stoichiometry of $A_{0.8}$Fe$_{1.6}$Se$_2$ (245) and shows a $\sqrt{5}\times\sqrt{5}$ Fe superstructure vacancy ordering \cite{Fang11,Bao11,Zhao12}. It exhibits a particular block antiferromagnetic state with a large magnetic moment per Fe atom of 3.3$ \mu_{\rm B}$ \cite{Bao11}. It has been argued that the minority 122-phase is metallic and becomes superconducting while the sample is cooled through the transition temperature \emph{T}$_{\rm c}$.  In contrary, the majority 245-phase with iron vacancy order remains magnetic, and hence the superconductivity occurs in the presence of the magnetic order.

Although, this phase separation scenario is established from the crystallographic point of view, the electronic nature of the different phases and their topological distribution is not yet clear.  Scanning photoelectron microscopy (SPEM) allows to spatially resolve the normally averaged spectroscopic information \cite{spectromicrosc}. In this work, for the first time, we have used SPEM technique to investigate the spatial distribution of the coexisting electronic phases in K$_{x}$Fe$_{2-y}$Se$_2$.

Figure \ref{Fig1}a shows a typical SPEM image for the angle-integrated photoelectron yield off the K$_{x}$Fe$_{2-y}$Se$_2$ at $T\simeq40$ K around the $\Gamma$-point. The photoemission intensity is integrated over a wide binding energy interval of $-1.5$ to 0.5 eV around the Fermi level $E_{\rm F}$, with a spatial resolution of $0.3\times0.3\,\mu$m$^2$ that reveals a clear inhomogeneous spectral distribution. The valence band spectra, measured around the $\Gamma$-point, on bright and dark regions are shown as Fig. \ref{Fig1}b. The bright region in the spectral image (Fig. \ref{Fig1}a) corresponds to the minority metallic phase showing a significant electronic density of states (DOS) at $E_{\rm F}$ (Fig. 1b).  On the other hand, dark region has hardly any DOS near $E_{\rm F}$ typical of  a semiconducting/insulating phase. Interestingly, the bright region is organized into interconnected thin filamentary stripes. 

In order to make a further characterization of the observed electronic phases, we have performed angle-resolved photoemission spectroscopy (ARPES)  measurements of the bright and dark regions to observe the band structure  in the $\Gamma$-M direction of the Brillouin zone.   The band dispersions for the bright and dark regions, using a photon energy  of 27 eV, are shown in Figs. \ref{Fig1}c and \ref{Fig1}d respectively. Despite limited energy and momentum resolutions ($\sim$ 50 meV and 0.05 $\AA^{-1}$), the energy dispersive structures in the bright regions of SPEM looks similar to the band structure measured by space integrated ARPES \cite{Zhang11,Qian11,Mou11,Liu12b,Chen12}. On the other hand, no dispersing structures are seen close to $E_{\rm F}$  in the dark region, neither around $\Gamma$ (zone center) nor around  M (zone corner), suggesting the dark region being due to an insulating/semiconducting phase. The difference between the two regions near the $E_{\rm F}$ can be better seen in Fig. \ref{Fig1}e, showing energy distribution of the spectral weight in the vicinity of the $E_{\rm F}$ obtained by integrating the band dispersions for the two regions (Figs. \ref{Fig1}c and \ref{Fig1}d). Therefore, we can make a clear distinction between the different electronic phases and their spatial distribution in the target material. Indeed, the bright filamentary stripes should represent the metallic and superconducting 122 phase, while the dark texture is the Fe vacancy ordered antiferromagnetic and semiconducting/insulating 245 phase.

The second derivative of the measured band dispersions for the bright filamentary stripes is shown in Fig. \ref{Fig2}a. Following earlier high resolution space integrated ARPES, the spectral weight around the zone center ($\Gamma$-point) can be assigned to weak electron-like band $\kappa$ crossing $E_{\rm F}$. This weak band is known to be mainly due to the Fe $3d_{xy}$ orbitals with admixed Se $4p_z$ orbitals \cite{Liu12b,Chen12}.  At slightly higher binding energies, around 80 meV below $E_{\rm F}$, spectral weight due to $\alpha$ and $\beta$ bands with hole-like character can be seen. These bands are mainly derived from a combination of the Fe $3d_{xz}$ and Fe $3d_{yz}$ orbitals.  At around 0.35 eV below $E_{\rm F}$, one can see the $\omega$ band, originating mainly from the Fe $3d_{3z^2}$ orbital. Around the zone corner (M point), spectral weight due to electron-like band $\delta$ can be identified that crosses $E_{\rm F}$ having a complicated orbital character. In addition, the $\beta$ band can be visualized also at the M point, around 0.2 eV below $E_{\rm F}$. The overall structures appear consistent with the band  dispersions reported  in the earlier space integrated ARPES \cite{Zhang11,Chen_PRX}.

To investigate energy dependence of the spatial distribution of the filamentary phase, we have measured SPEM images with different binding energy windows identified from the band dispersions. Figs. \ref{Fig2}b to \ref{Fig2}g show the spectral intensity distributions for the different bands.  Figure \ref{Fig2}b represents spectral intensity distribution of the electron-like $\kappa$ band close to $E_{\rm F}$ ($-0.07$~to~0 eV below $E_{\rm F}$).  From the band-selected SPEM images (in particular, those for the $\kappa$ and $\alpha$/$\beta$ bands around the $\Gamma$-point),  it is clearly seen that the spatial phase separation  in K$_{x}$Fe$_{2-y}$Se$_2$ occurs with interconnected thin filamentary stripes of the superconducting 122 phase organized in a peculiar geometry.  The filamentary stripes have a width of less than 1 $\mu$m, and are embedded into the dark region (i.e., the semiconducting/insulating 245 phase). 

From the spectral intensity map, the ratio of the 245 phase with lower DOS  to the 122 phase with higher DOS can be determined. A detailed analysis of the distribution histogram of the $\alpha$/$\beta$  bands shown in the inset of Fig. \ref{Fig2}c reveals an asymmetric intensity distribution which can be fitted to a sum of two Gaussians. This analysis  indicates that $\simeq85$\% of the sample can be attributed  to the insulating/semiconducting  245 and the remaining $\simeq15$\% to the metallic 122 phase.  This is consistent with the earlier estimations of the volume fractions of the majority and minority phases determined by different experiments \cite{Shermadini12,Ricci11b,Wen12}. The SPEM images underline that instead of granular superconductivity a scenario of filamentary superconductivity is more plausible for K$_{x}$Fe$_{2-y}$Se$_2$. In addition, the SPEM image of the $\alpha$/$\beta$ bands around $\Gamma$ shows  very similar features as the $\kappa$ band around $\Gamma$ (see Fig.  \ref{Fig2}c).  Also the $\omega$ band shows (Fig. \ref{Fig2}d) network of filaments of the 122 phase.  As expected a very similar spatial distribution of the spectral weight of the bands around the M point, where the $\delta$ band crossing $E_{\rm F}$, is observed (Fig. \ref{Fig2}e). The same is observed for the $\beta$ band at M (see Fig.  \ref{Fig2}f).   In comparison to the energy integrated SPEM image presented in Fig.~\ref{Fig1}b, the SPEM images with energy and momentum selected windows show significantly higher contrast resolution in the distribution of the filamentary metallic 122 phase. 

At higher binding energies, the SPEM images appear quite uniform (Fig.  \ref{Fig2}g), however, still some inhomogeneities can be seen. The corresponding intensity histogram is symmetric and can be fitted by a single Gaussian (see the inset to Fig. \ref{Fig2}g). Nevertheless, the SPEM images permit to conclude unambiguously that the peculiar electronic phase separation in K$_{x}$Fe$_{2-y}$Se$_2$ is solely due to the bands close to $E_{\rm F}$ with Fe 3$d$ character. 

Here, it is worth mentioning that a structural phase separation was observed, with uniform areas of the 245 phase enclosed by a network of the 122 phase  having a width of about 1 $\mu$m \cite{Charnukha12,Liu12, Wen12}. Also a combination of scanning and transmission electron microscopy has shown a structural phase separation in this scale \cite{Wang12}. The peculiar  topological distribution of the metallic phase in K$_{x}$Fe$_{2-y}$Se$_2$ is characterstic of the system. The distribution of the filamentary phase may be different depending on the sample preparation procedure and thermal history. Indeed, the contrast resolution and topological distribution of coexisting phases appear to depend on the sample heat treatment, also underlined by transmission electron microscopy \cite{Wang12}, measureing different structural phases unlike the present work in which the electronic structure of the coexisting phases is determined.

Clearly, the electronic phase separation is characterized by Fe $3d$ electronic states.  In order to have a further insight to the role of other elements, core level SPEM of K and Se were performed. Figure  \ref{Fig_corelevel}a shows the spatial distribution of the integrated photoemission yield measured using a photon energy of 74 eV  around the K 3$p$ core level. No clear inhomogeneities can be observed in the SPEM image, indicating an overall homogeneous distribution of K.  However, a more detailed investigation leads to the observation of the binding energy shift in the K 3$p$ peak. The inset to Fig. \ref{Fig_corelevel}a shows the  K 3$p$ photoemission peaks of the metallic 122  (bright) and the insulating/semiconducting 245 (dark) phases. The K 3$p$ binding energy for the 245 phase is lower than the one for the 122 phase, indicating stronger screening. It appears that  the valence of K is different for the two phases, perhaps due to different interactions with the neighbouring atoms (e.g., degree of covalency). Also, the K 3$p$ core level appears wider for the 122 phase, which might be due to  mixing of the 245 phase in the 122 phase due to the fact that  the filaments of the latter are less than 1$\mu$m wide. Similarly, the spatial distribution of Se 3$d$ core level photoemission intensity was measured.  Although the Se 3$d$ peak is overlapping with the Fe 3$p$ core level, the lower cross section of Fe 3$p$ compared to the Se 3$d$ at the photon energy used allowed to measure the desired SPEM image.  The resulting map of the Se $3d$ shows no evident inhomogeneities (see Figs.  \ref{Fig_corelevel}b).  Thus, it appears that, similar to K, Se is also distributed uniformly in the sample. Within the experimental resolution, no evidence of Se 3$d$ core level shift was found, suggesting almost similar oxidation state for Se in the two phases.

From the SPEM imaging, the electronic phase separation is clearly observed in K$_{x}$Fe$_{2-y}$Se$_{2}$.  The sample consists predominantly of the insulating 245 phase that is interrupted by filamentary interconnected stripes with less than 1 $\mu$m size of 122 phase.  It seems that K and Se are distributed homogeneously in the sample leading to the conclusion that mainly the Fe content $y$ is varying due to the phase separation into the stoichiometric 122 and the Fe deficient 245 phase.  That could indicate that $x$ and $y$ are strongly correlated and solely the Fe content $y$ is responsible for the electronic phase separation.  Consequently, Fe might posses different spin states in the corresponding electronic phases.  Such a scenario was already observed in X-ray emission studies with a majority phase having a high spin Fe$^{2+}$ and a minority phase having an intermediate Fe$^{2+}$ state \cite{Simonelli12}.   This appears consistent with the observation of different covalance of K depending on the corresponding phase as observed in the shift of the K 3$p$ core level SPEM.  The present results may have implications on the recent indications of superconductivity in a single layer of FeSe grown on SrTiO$_{3}$ substrate \cite{WangFeSe12}, where electronic structure studies have shown absence of hole bands \cite{LiuFeSe12,TanFeSe13} as the case of K$_{x}$Fe$_{2-y}$Se$_{2}$.  Indeed, a particular filamentary network can give rise to high T$_{c}$ superconductivity as argued in the oxide interfaces \cite{Caprara13}. From the SPEM results, it can be hypothesized that a particular electronic structure topology of the conducting phase (FeSe) embedded in an insulating texture (SrTiO$_{3}$ substrate) is a key to the higher T$_{c}$ superconductivity in the single layer FeSe.

\section*{Methods}
{\footnotesize 
\textbf{Sample growth} The K$_x$Fe$_{2-y}$Se$_2$ single crystals were prepared using the Bridgman method
\cite{Mizuguchi11}.  After the growth the single crystals were sealed into a quartz tube and annealed for 12 hours in 600$^\circ$C.  Some of the crystals were annealed in vacuum at 400$^\circ$C for 1 hour  followed by fast cooling to room temperature.  The crystals were characterized by X-ray diffraction, resistivity in a PPMS (Quantum Design), and magnetization measurements in a SQUID (Quantum Design) magnetometer. The samples exhibit a sharp superconducting transition with $T_{\rm c}\simeq 32$ K. \\ \textbf{Spectromicroscopy.} The SPEM measurements were performed on the spectromicroscopy beamline at the 'Elettra synchrotron facility'  at Trieste \cite{spectromicrosc}.  Photons at 27 eV and 74 eV were focused through a Schwarzschild objective, to obtain a submicron size spot.  This allows to sample the electronic structure at the Fermi level as well as several core levels along with the valence level in order to visualize the normally averaged spectroscopic information in the electronic structure.  For the present measurements the total energy resolution was about 50 meV while the angle resolution was limited to 1$^\circ$. The measurements were performed in ultra-high vacuum (p $<$ 2 $\times $10$^{- 10}$ mbar) on an in-situ prepared (cleaved) surfaces.  The SPEM images contain 150x144 points with a step of 0.3 $\mu m$ and the measurement time for each point is 100 milliseconds. A standard photoemission microscopy procedure was used to remove topographic features from the images presented \cite{surface}.}

{\footnotesize
\section*{Acknowledgements}
The authors would like to thank Elettra staff for the experimental help during the beamtime. This work is a part of the on-going collaboration between Sapienza University of Rome and Japanese institutes including University of Tokyo, NIMS and Tohoku University. D.I.
and A. Bi. acknowledge support from superstripes – onlus. M. B. acknowledges support from the Swiss National Science Foundation (grant number PBZHP2\_143495).

\section*{Author contributions}
N.L.S.,T.M., M.B., A.F., Y.T. Y.K, A.Ba. and A.Bi programmed and coordinated the study. H. T., Y. M., Y. T., T. N., T. H. and Y. K. have synthesized the single crystals used for the study. M.B., N.L.S., A.Ba., M.H., T.M., A.I., D.I. and D.O.  performed the experiments at  the Spectromicroscopy beamline of Elettra synchrotron facility and contributed in the data analysis. M.B., B.J. and N.L.S wrote the paper with contributions from A.Ba. and T.M. All authors discussed the results and commented on the manuscript.

\section*{Additional information}
The authors declare no competing financial interests. Reprints and permissions information is available online at http://www.nature.com/reprints. 
Correspondence and requests for materials should be addressed to N.L.S. (Naurang.saini@roma1.infn.it)

\begin{figure}[t]
\centering
\vspace{-0cm}
\hspace{0cm}\includegraphics[width=\linewidth]{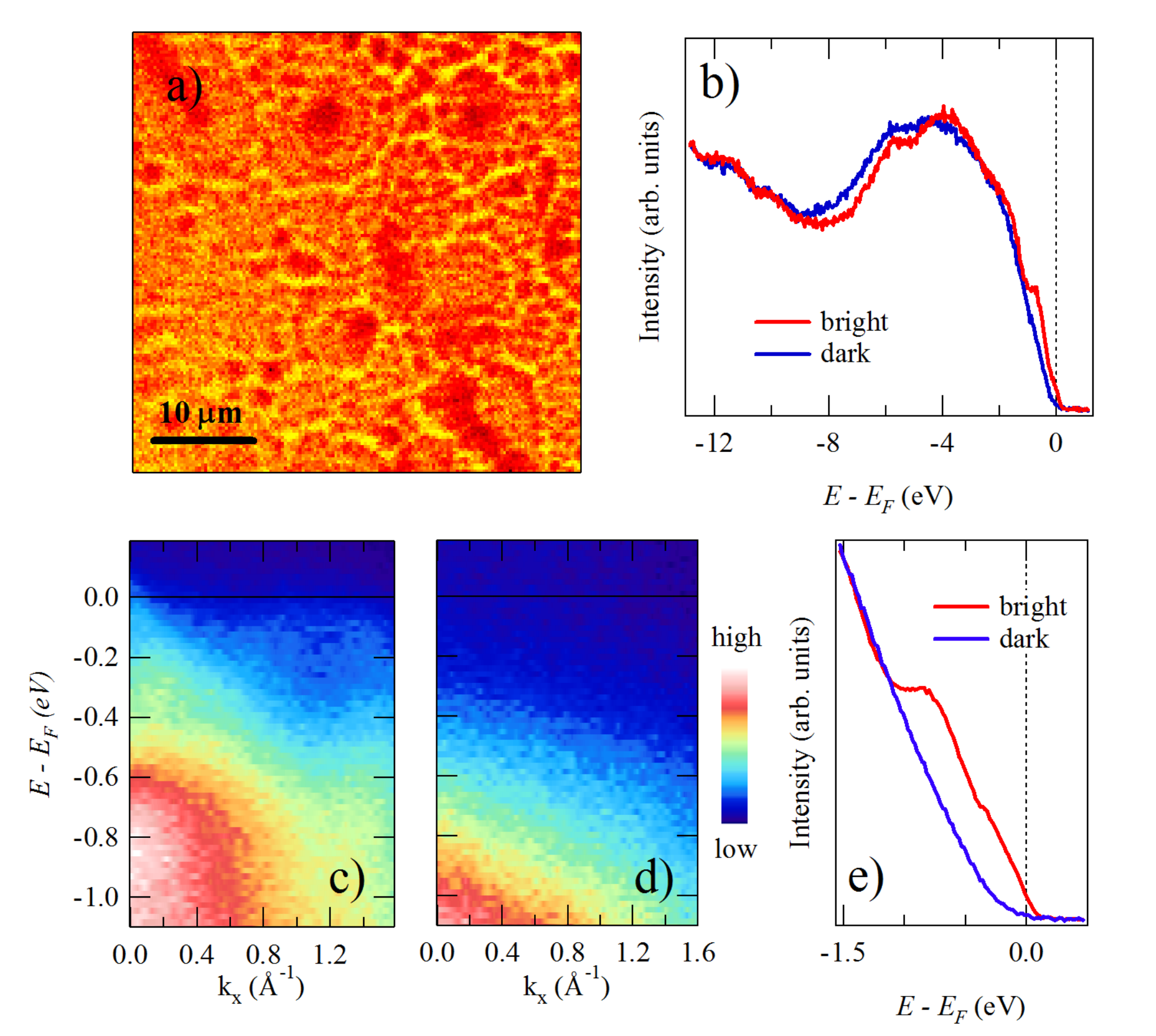}
\vspace{-0.5cm}

\caption{\textbf{Scanning photoelectron microscopy images and the electronic structure of K$_{1-x}$Fe$_{2-y}$Se$_{2}$ at 40 K.  a,} Overview image with the spatial resolution given by the pixel size of $0.3\times0.3$ $\mu$m$^2$.  With this resolution, one can see a peculiar topological distribution of an electronic phase with higher DOS. \textbf{b} Valence band spectra of the bright and dark regions near $\Gamma$-point.
Band dispersions along the $\Gamma$-M of the bright region \textbf{(c)}  and dark region \textbf{(d)}. Energy distribution of the spectral weight near the $E_{\rm F}$ is also shown (\textbf{e}), that makes a clear distinction between the dark region (lower DOS) and the bright region (higher DOS).} 
\label{Fig1}
\end{figure}

\begin{figure*}[t]
\centering
\vspace{-0cm}
\hspace{0cm}\includegraphics[width=\linewidth]{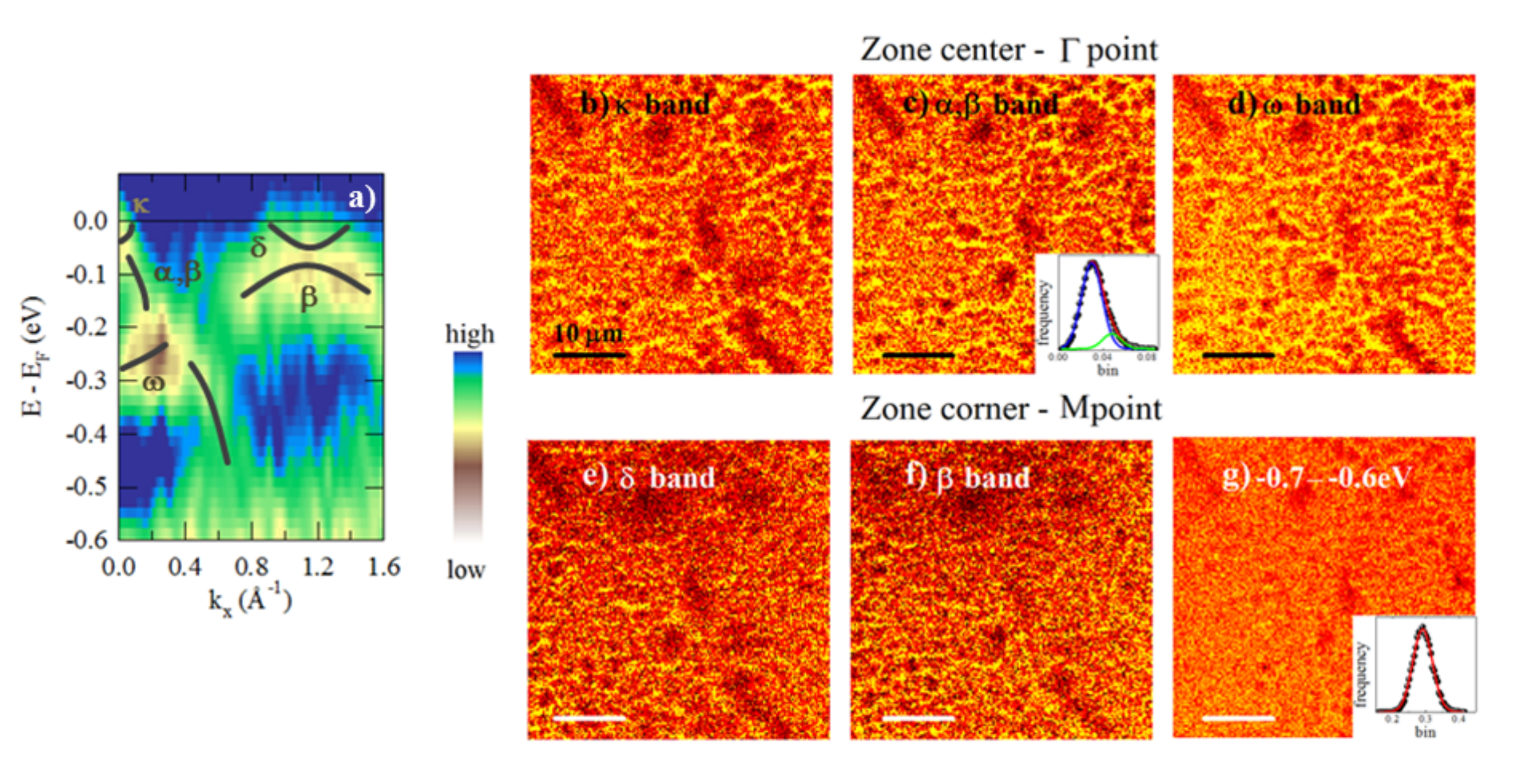}
\vspace{-0.5cm}
\caption{\textbf{SPEM for the specific electronic bands at different binding energies in K$_{1-x}$Fe$_{2-y}$Se$_{2}$. a,} Second derivative of the measured band dispersion of the bright region along the $\Gamma$-M direction. \textbf{b,} Spatial distribution of the photoelectron yield for the $\kappa$ band  at the center of the Brillouin zone ($\Gamma$-point). The bright region represents the paramagnetic and metallic 122 filamentary phase embedded into the matrix of the antiferromagnetic and semiconducting/insulating 245 phase. \textbf{c,}  Spatial distribution of the photoelectron intensity of the $\alpha$ and $\beta$ bands around the $\Gamma$ point.  The inset shows the asymmetric intensity histogram  that is described by a sum of two Gaussian functions.  \textbf{d,} Spatial distribution of the photoelectron intensity for the $\omega$ band.  The interconnected filamentary network is still present, however, less pronounced. \textbf{e,} Spatial distribution of the photoelectron intensity for the $\delta$ band. The image exhibits the similar features. \textbf{f,} Spatial distribution of the photoelectron intensity for the $\beta$ band around the M point, revealing the similar features. \textbf{g,} Spatial distribution of the photoelectron intensity for the higher energy region ($-0.7$ to $-0.6$ eV below $E_{\rm F}$). The electronic structure seems more uniform with small inhomogeneities.  The inset shows the intensity histogram described by a Gaussian function.} \label{Fig2}
\end{figure*}

\begin{figure}[tb]
\centering
\vspace{-0cm}
\hspace{0cm}\includegraphics[width=\linewidth]{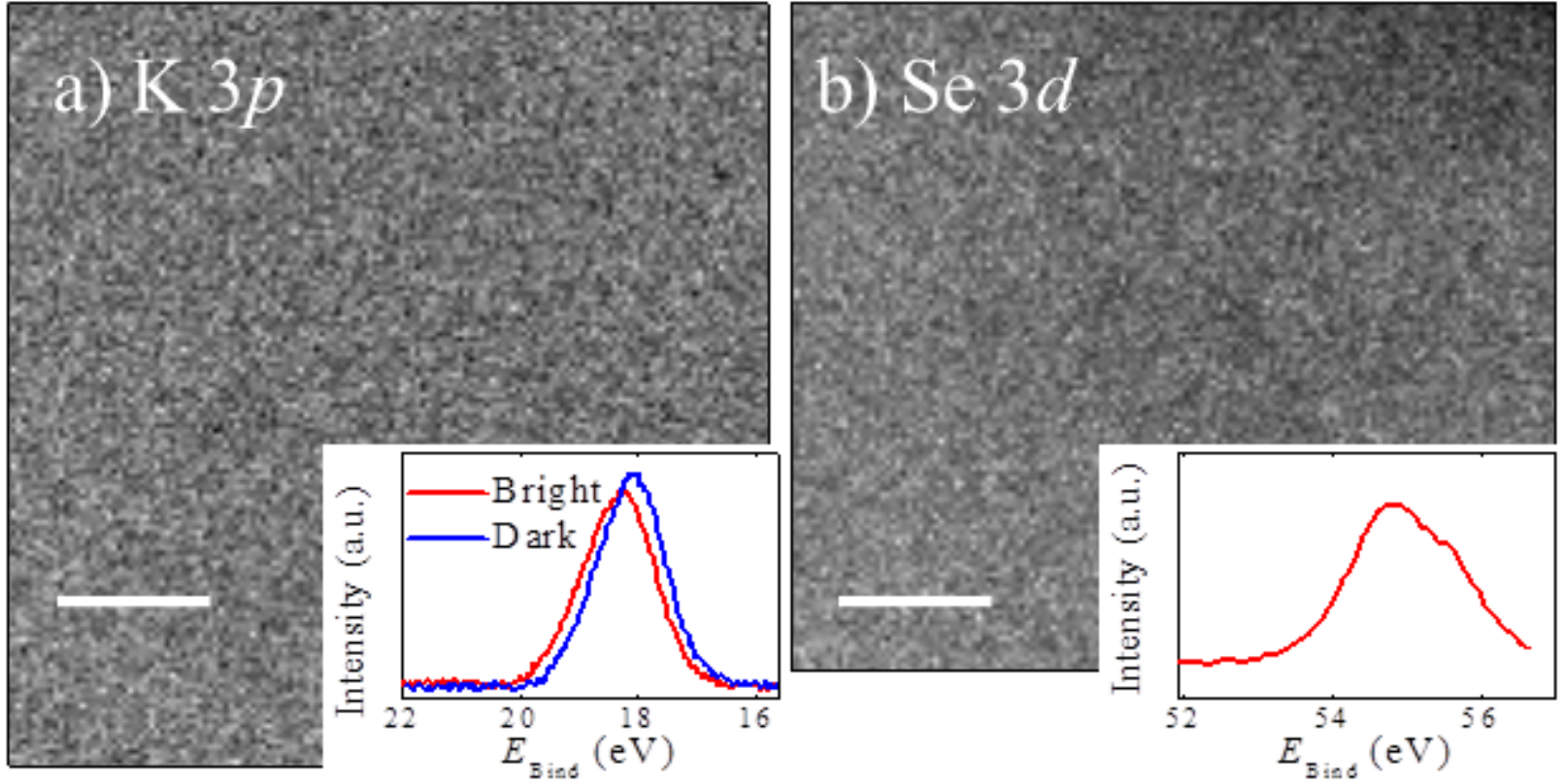}
\vspace{-0.5cm}
\caption{\textbf{Core-level SPEM images of K$_{1-x}$Fe$_{2-y}$Se$_{2}$. a,}  Spatial distribution of K 3$p$ spectral weight, indicating a homogeneous  K distribution.  The inset shows the K 3$p$ core level photoemission spectra for the 245 phase (blue) and the 122 phase (red).  The shift reveals different binding energies in the corresponding phases. \textbf{b,} Se 3$d$ core level SPEM revealing a rather uniform distribution of Se.  The inset shows Se 3$d$ core level photoemission spectra.  The white scale bars in the figures represent
$10~\mu$m.} \label{Fig_corelevel}
\end{figure}

\end{document}